# Extraordinary cation-replace-cation antisite defect predominate in $Bi_2SeO_5$

Chen-Min Dai,[1,2] Feifan Bian,[1] Yafeng Zhang,[1] Jiaqi Chen,[1] Zenghua Cai,[1,2*] Menglin Huang,[3*] Chunlan Ma,[1,2*]

[1]Jiangsu Key Laboratory of Intelligent Optoelectronic Devices and Chips, School of Physical Science and Technology, Suzhou University of Science and Technology, Suzhou, 215009, China

[2]Advanced Technology Research Institute of Taihu Photon Center, School of Physical Science and Technology, Suzhou University of Science and Technology, Suzhou, 215009, China

[3]College of Integrated Circuits and Micro-Nano Electronics, and Key Laboratory of Computational Physical Sciences (MOE), Fudan University, Shanghai, 200433, China

*Contact author:zhcai@usts.edu.cn;menglinhuang@fudan.edu.cn;wlxmcl@usts.edu.cn

## Abstract

As a newly identified single-crystalline van der Waals dielectric with a high dielectric constant, $Bi_2SeO_5$ plays a pivotal role in advancing 2D electronic devices. In this work, we systematically investigate the defect properties of $Bi_2SeO_5$ using first-principles calculations based on a hybrid functional. Although $Bi_2SeO_5$ is a chemically ternary compound, each constituent element occupies several crystallographically nonequivalent sites, rendering its defect chemistry highly complex. Due to the anomalous +4 cationic valence state of Se, the defect formation energies of same main group anion antisite defects ($Se_O$ and $O_{Se}$) are prohibitively high, and their concentrations can therefore be neglected. In contrast, the extraordinary cation–cation antisite defects $Bi_{Se}$ and $Se_{Bi}$ emerge as the dominant defects. The pronounced variability in the formation energies of the six types of $V_O$ defects demonstrates that identical defect types located on nonequivalent atomic sites can exhibit markedly different properties. Under O-rich and Se/Bi-poor conditions, $Bi_2SeO_5$ shows relatively robust p-type behavior. Conversely, under O-poor and Se/Bi-rich conditions, or at intermediate O, Se, and Bi partial pressures, $Bi_2SeO_5$ behaves as an intrinsic semiconductor or displays very weak n-type conductivity due to strong donor–acceptor compensation. This study provides theoretical insights to guide the design and development of high-performance $Bi_2SeO_5$-based electronic devices.

## Introduction

Silicon-based semiconductor technology is increasingly constrained by intrinsic material limits and fabrication challenges at sub-7 nm nodes. Conventional dielectrics such as $SiO_2$ and $HfO_2$ often introduce interfacial defects due to nucleation, reducing their compatibility with pristine surfaces. Against this backdrop, two-dimensional (2D) materials have attracted significant attention as promising candidates for next-generation nanoelectronics. Their atomically thin structures and absence of dangling bonds enable stable gate modulation in ultra-scaled channels. Hexagonal boron nitride, the most widely adopted van der Waals dielectric, suffers from a low dielectric constant, resulting in excessive gate leakage and poor electrostatic control when thinned[1]. Despite advances in high-mobility channel materials, progress has been consistently limited by the absence of van der Waals dielectrics that simultaneously offer a high dielectric constant and minimal interfacial defects[2-4]. These limitations underscore the urgent need for novel van der Waals dielectrics that combine high permittivity with low defect density.

The layered dielectric $Bi_2SeO_5$ (BSO) has emerged as a promising candidate for advanced electronic applications owing to its exceptional properties[5-7]: including a wide band gap, high dielectric constant, and excellent carrier mobility. Recent breakthroughs have demonstrated the fabrication of atomically thin BSO as a gate dielectric in field-effect transistors (FETs), where $Bi_2O_2Se$/BSO FETs outperform conventional $HfO_2$-based devices[8]. These devices exhibit outstanding characteristics—such as a low subthreshold swing, negligible hysteresis, high mobility, and an impressive current on/off ratio-arising from the crystalline nature of BSO, which effectively minimizes oxide trap density. The synergy between a high-mobility 2D semiconductor $Bi_2O_2Se$ and its native oxide BSO positions this system as a strong contender for post-Si/$SiO_2$ electronics[8-10].

Beyond transistors, the unique crystal and electronic structures of BSO have enabled diverse applications[11-15], including photodetectors, memristors, sensors, batteries, and more. Oxygen-vacancy-engineered BSO memristors exhibit superior nonvolatile switching behavior, with conductive filament formation and rupture mechanisms experimentally confirmed[12]. Defect-introduced BSO nanosheets enhance $NH_4^+$ battery performance by weakening electrostatic interactions and accelerating $NH_4^+$ diffusion. Oxygen vacancies simultaneously create active sites, facilitate electron transfer, and improve conductivity[15]. Experimental evidence consistently demonstrates that both intrinsic defects and engineered defect configurations in BSO can significantly enhance device performance. Owing to its two-dimensional crystal characteristics, BSO inherently mitigates interfacial defect states, while intrinsic defect properties critically influence the material's transport behavior. Although theoretical studies have explored the dielectric[16, 17], mechanical[18], and optical properties[19] of BSO, systematic investigations of point defects remain scarce, yet they hold considerable potential for unlocking further performance improvements.

In this work, we systematically study the crystal structure, thermodynamic stability, and defect properties of BSO using first-principles calculations based on a hybrid functional. In the BSO, the complex atomic bonding environment gives rise to numerous crystallographically nonequivalent positions for different atomic species, rendering its defect properties more intricate than those of simple ternary compounds. Although the stable chemical potential region of pure BSO is relatively wide, careful control of the chemical potential is required to avoid impurities phases such as $Bi_2Se_3O_9$, $Bi_2Se_3O_{10}$ and $Bi_2O_3$. The defect behavior of BSO proves to be highly unconventional. Defect calculations reveal that $Bi_{Se}$ and $Se_{Bi}$ are the dominated deep acceptor and donor defects, respectively, with concentrations exceeding those of $V_O$, which chemical intuition would predict to form more readily. In traditional compounds, Se typically exhibits anionic characteristics with a -2 valence state, whereas in BSO it adopts cationic +4 state. This unusual chemistry facilitates the formation of a high density of cation-cation antisite defects $Bi_{Se}$ and $Se_{Bi}$ , which emerge as the predominant defects. Furthermore, $V_O$ defects occur in six distinct configurations, each exhibiting markedly different defect properties. Under O-rich and Se/Bi-poor conditions, BSO displays relatively robust p-type behavior. In contrast, under O-poor and Se/Bi-rich conditions, or at intermediate O, Se, and Bi partial pressures, BSO behaves as an intrinsic semiconductor or exhibits very weak n-type conductivity due to strong donor-acceptor compensation.

## Calculation Methods

The first-principles calculations were performed using the density functional theory as implemented in the Vienna *ab initio* simulation package (VASP) code[20]. The projector augmented-wave (PAW)[21, 22] pseudopotentials with an energy cutoff of 600 eV for the plane-wave basis set were employed. All lattice vectors and atomic positions were relaxed until the total force on each atom was lower than 0.01 eV/Å. The Heyd-Scuseria-Ernzerhof (HSE06)[23, 24] hybrid functional with an exchange parameter of 0.18 was utilized in total energy calculation for each defect. A single Γ point mesh[25, 26] was used for all the defect simulations based on the supercell with 128 atoms.

In order to fabricate the pure BSO, the chemical potential of the composition elements should be limited by a series of thermodynamic conditions. Firstly, the chemical potential of Bi, O, and Se in stable BSO compound must satisfy the following equation.

$$2\mu_{\mathrm{Bi}} + \mu_{\mathrm{Se}} + 5\mu_{\mathrm{O}} = \Delta H_f\left(\mathrm{Bi_2SeO_5}\right) = \text{-11.82 eV},$$

where $\mu_i$ represent the chemical potential of element $i$ (Bi, O, Se), is the calculated formation energy of BSO. Secondly, to avoid the competing phase (Bi, $O_2$, Se, $Bi_2O_3$, $Bi_2Se_3$, $Bi_4O_7$, $Bi_4Se_3$, $BiO_2$, BiSe, $BiSe_2$, $Se_2O_5$, $SeO_2$, $Bi_2O_2Se$, $Bi_2Se_3O_{10}$ and $Bi_2Se_3O_9$) forming, several inequality relations should be required.

$$\mu_{\mathrm{Bi}} < 0, \quad \mu_{\mathrm{Se}} < 0, \quad \mu_{\mathrm{O}} < 0$$

$$2\mu_{\mathrm{Bi}} + 3\mu_{\mathrm{O}} < \Delta H_{\mathrm{f}}(\mathrm{Bi_2O_3}) = \text{-7.58 eV},$$

$$2\mu_{\mathrm{Bi}} + 3\mu_{\mathrm{Se}} < \Delta H_{\mathrm{f}}(\mathrm{Bi_2Se_3}) = \text{-1.92 eV},$$

$$4\mu_{\mathrm{Bi}} + 7\mu_{\mathrm{O}} < \Delta H_{\mathrm{f}}(\mathrm{Bi_4O_7}) = \text{-15.73 eV},$$

$$4\mu_{\mathrm{Bi}} + 3\mu_{\mathrm{Se}} < \Delta H_{\mathrm{f}}(\mathrm{Bi_4Se_3}) = \text{-1.87 eV},$$

$$\mu_{\mathrm{Bi}} + 2\mu_{\mathrm{O}} < \Delta H_{\mathrm{f}}(\mathrm{BiO_2}) = \text{-3.96 eV},$$

$$\mu_{\mathrm{Bi}} + \mu_{\mathrm{Se}} < \Delta H_{\mathrm{f}}(\mathrm{BiSe}) = \text{-0.62 eV},$$

$$\mu_{\mathrm{Bi}} + 2\mu_{\mathrm{Se}} < \Delta H_{\mathrm{f}}(\mathrm{BiSe_2}) = \text{-0.86 eV},$$

$$2\mu_{\mathrm{Se}} + 5\mu_{\mathrm{O}} < \Delta H_{\mathrm{f}}(\mathrm{Se_2O_5}) = \text{-6.35 eV},$$

$$\mu_{\mathrm{Se}} + 2\mu_{\mathrm{O}} < \Delta H_{\mathrm{f}}(\mathrm{SeO_2}) = \text{-3.13 eV},$$

$$2\mu_{\mathrm{Bi}} + 2\mu_{\mathrm{O}} + \mu_{\mathrm{Se}} < \Delta H_{f}(\mathrm{Bi_2O_2Se}) = \text{-5.88 eV},$$

$$2\mu_{\mathrm{Bi}} + 3\mu_{\mathrm{Se}} + 10\mu_{\mathrm{O}} < \Delta H_{f}(\mathrm{Bi_2Se_3O_{10}}) = \text{-19.94 eV},$$

$$2\mu_{\mathrm{Bi}} + 3\mu_{\mathrm{Se}} + 9\mu_{\mathrm{O}} < \Delta H_{f}(\mathrm{Bi_2Se_3O_9}) = \text{-19.06 eV}.$$

The formation energy $\Delta E_f(\alpha,q)$ of a defect $\alpha$ in the charge state $q$ is calculated by[27-29]

$$\Delta E_f(\alpha,q) = E(\alpha,q) - E(host) + \sum_i n_i(E_i+\mu_i) + q\left[\epsilon_{\mathrm{VBM}}(host) + E_{\mathrm{F}} + \Delta V_{align}\right] + E_{corr}$$

where $E(\alpha,q)$ is the total energy of the supercell with defect $\alpha$ and $E(host)$ is the total energy of the host, $\mu_i$ is the atomic chemical potential of constituent $i$ (Bi, O, Se) referenced to the total energy $E_i$ of its pure elemental solid. $n_i$ is the number of atom $i$ and $q$ is the number of electrons exchanged between the supercell and the corresponding thermodynamic reservoir when the defect forms. $\epsilon_{\mathrm{VBM}}(host)$ is the valence band maximum (VBM) referring to the Fermi energy level, and $E_{\mathrm{F}}$ is the Fermi energy level. $\Delta V_{align}$ aligns the electrostatic potential far from the defect site in the defect and host supercells. We adopt Freysoldt-Neugubauer-Van de Walle scheme[30] for finite-size correction for charged defects.

## Nonequivalent Atomic Positions

BSO crystallizes in the *Abm2* space group with lattice constants $a$ = 11.42 Å, $b$ = 16.24 Å, $c$ = 5.49 Å. As illustrated in Figure 1, the BSO primitive cell contains 32 atoms: 8 Bi, 4 Se, and 20 O. Among these atoms, Bi occupies three distinct sites (Bi1, Bi2, Bi3), O occupies six distinct sites (O1–O6) and Se occupies a single site. Bi1 is coordinated by eight oxygen atoms, whereas Bi2 and Bi3 are each bonded to seven oxygen atoms. Among the oxygen atoms, O1, O3, and O4 are connected to both Bi and Se neighbors, while O2, O5, and O6 bond exclusively to Bi. These variations in coordination create unique chemical environments at each site, leading to diverse defect characteristics. Consequently, point defects originating from these nonequivalent positions are expected to exhibit significant differences in both formation energies and electronic properties.

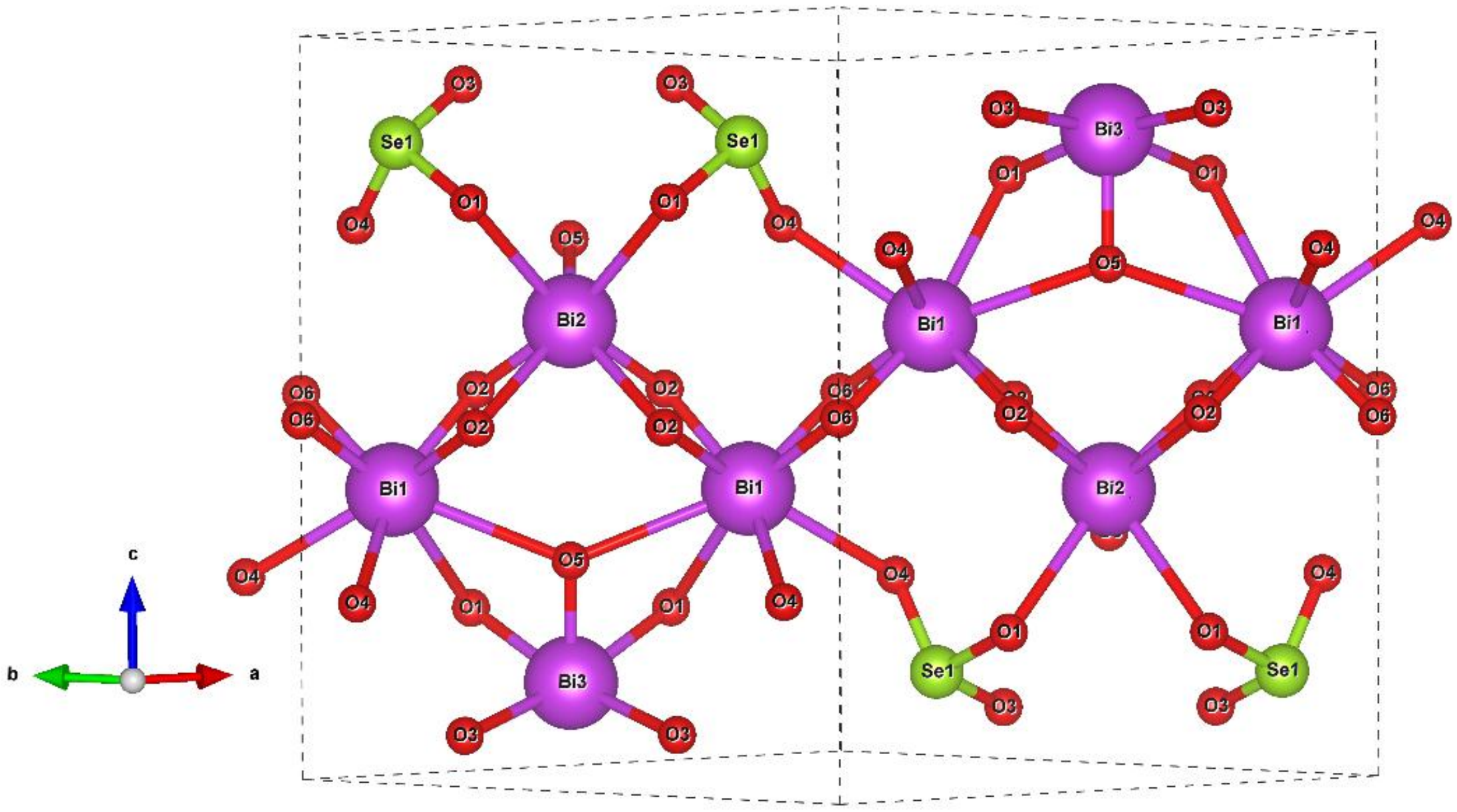

Figure 1. Nonequivalent atomic positions in the primitive cell of $Bi_2SeO_5$.

We comprehensively considered all intrinsic point defects in BSO, including six oxygen vacancies ($V_{O1}$-$V_{O6}$), six bismuth antisites on oxygen sites ($Bi_{O1}$-$Bi_{O6}$), six selenium antisites on oxygen sites ($Se_{O1}$-$Se_{O6}$), three bismuth vacancies ($V_{Bi1}$-$V_{Bi3}$), three oxygen antisites on bismuth sites ($O_{Bi1}$-$O_{Bi3}$), three selenium antisites on bismuth sites ($Se_{Bi1}$-$Se_{Bi3}$), one selenium vacancy ($V_{Se1}$), one oxygen antisite on selenium sites ($O_{Se1}$), and one bismuth antisite on selenium sites ($Bi_{Se1}$). For each interstitial species ($Bi_i$, $O_i$, $Se_i$), twenty candidate configurations are generated, and the lowest-energy structure was selected for defect-formation analysis.

## Stable Thermodynamic Region of $Bi_2SeO_5$

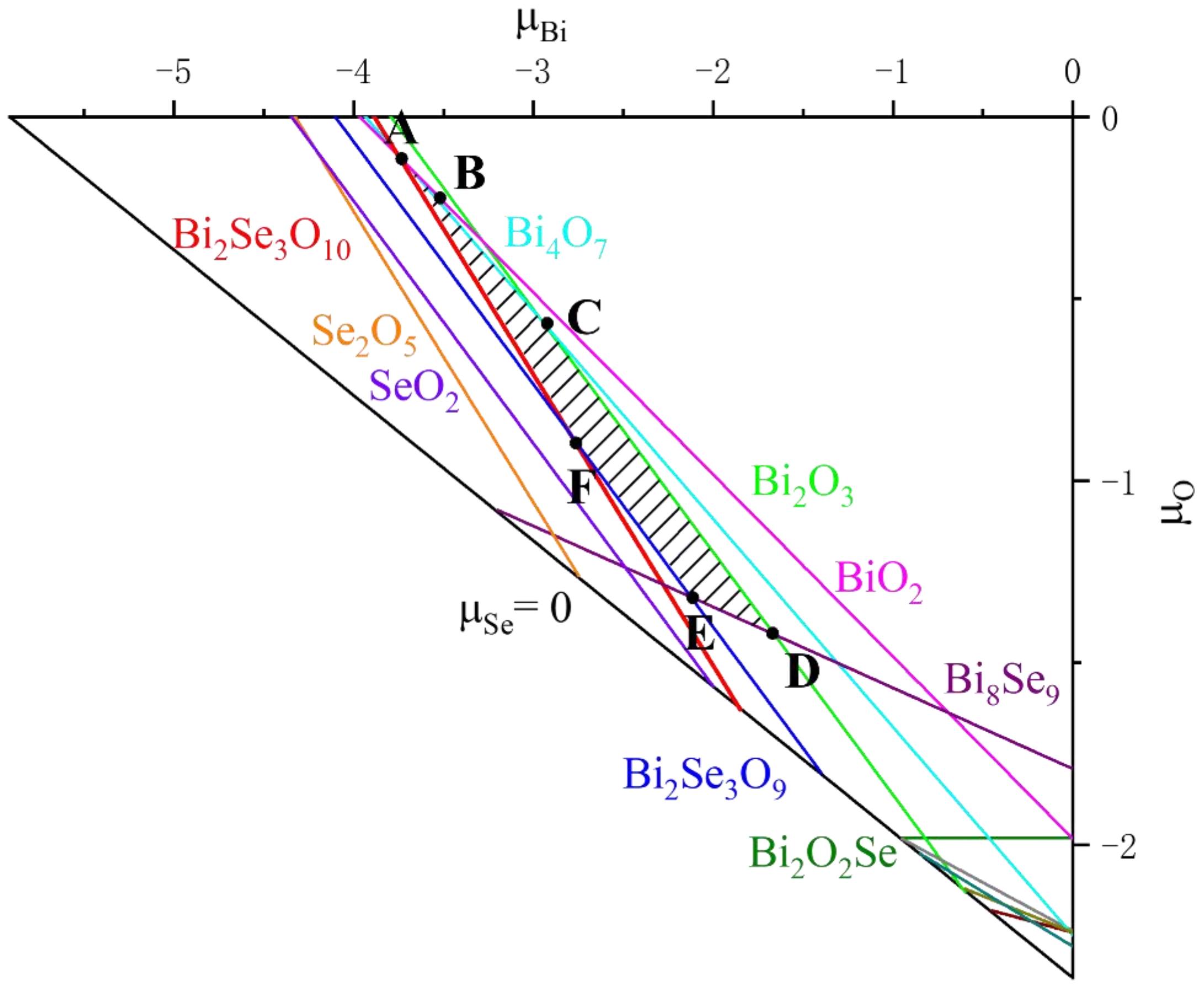

Figure 2. Projection of the chemical potential stability region of $Bi_2SeO_5$ in the ($\mu_O$, $\mu_{Bi}$) plane. The black shaded region indicates the chemical potentials where single-phase $Bi_2SeO_5$ forms. Points A ($\mu_{Bi}$ = -3.76 eV, $\mu_{Se}$ = -3.9 eV, $\mu_O$ = -0.10 eV), B (-2.71, -3.77, -0.13), C (-3.12, -2.95, -0.56), D (-1.66, -1.39, -1.42), E (-2.12, -0.98, -1.32) and F (-2.8, -1.88, -0.87) represent six typical growth conditions.

As shown in Figure 2, BSO is thermodynamically stable when the chemical potentials of the elements fall within the black shaded region. Beyond this region, secondary phases inevitably coexist. Under increasingly oxygen-poor conditions, $BiO_2$, $Bi_4O_7$, and $Bi_2O_3$ emerge as the primary competing oxides, whereas elevated selenium potentials favor the formation of $Bi_8Se_9$, $Bi_2Se_3O_{10}$, and $Bi_2Se_3O_9$. Thus, the synthesis of pure BSO requires precise control of Bi, Se, and O chemical potentials. Points A - F represent six distinct atomic chemical potentials (growth conditions), which are critical factors for regulating defect properties. The ranges of $\mu_{Bi}$, $\mu_{Se}$, and $\mu_O$ span -3.76 to -1.66 eV, -3.9 to -0.98 eV, and -1.42 to -0.1 eV, respectively.

## Defect Properties of Nonequivalent Atomic Positions

Figure 3 illustrates the formation energies for $V_O$, $V_{Bi}$, and $Se_{Bi}$ as functions of the Fermi level under chemical potential condition (point A). All nonequivalent atomic positions were systematically evaluated. The non-overlapping formation energy curves for identical defect types at different sites reveal distinct defect behaviors at nonequivalent atomic positions.

Figure 3(a) highlights six distinct $V_O$ configurations. Within the energy window from the conduction band minimum (CBM) to 1.5 eV below, all $V_O$ defects remain in neutral charge states, with formation energies independent of the Fermi level. While $V_{O1}$ and $V_{O3}$ exhibit nearly identical neutral formation energies, the other configurations show pronounced disparities. The formation energies of $V_{O1}$, $V_{O3}$, and $V_{O4}$ are lower than those of $V_{O2}$, $V_{O5}$, and $V_{O6}$. Notably, a 0.5 eV difference of formation energy corresponds to an approximately 8 orders of magnitude variation in defect concentration, and the neutral formation energies of $V_{O4}$ and $V_{O2}$ differ by as much as 1.2 eV. As the Fermi level shifts downward, each $V_O$ defect ionizes sequentially, and their formation energies become highly sensitive to the Fermi level. The +2 charged state ($V_O^{2+}$) emerges as the thermodynamically preferred configuration due to its reduced formation energy. When the Fermi level approaches the mid-gap position, $V_{O3}^{2+}$ has the lowest formation energy, whereas $V_{O1}^{2+}$ reaches the highest, separated by 0.5 eV (as shown in Fig. 3b)—an energy scale sufficient to profoundly influence defect concentrations.

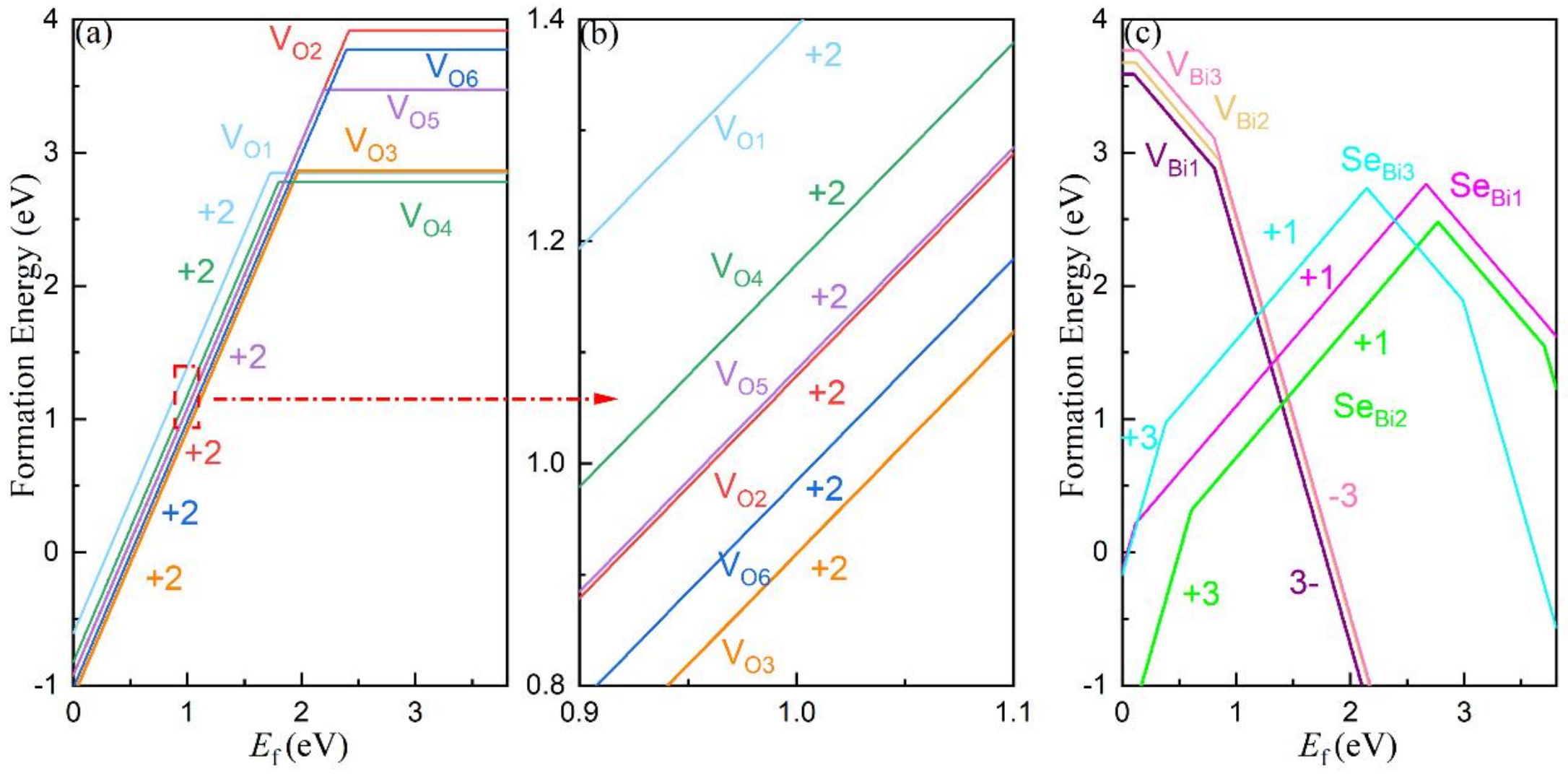

Figure 3. (a) Defect formation energies of $V_O$, (c) $V_{Bi}$, and $Se_{Bi}$ at different atomic sites as functions of the Fermi level. (b) is an enlarged view of the $V_O$ defect region from (a).

All six $V_O$ defects act as donors, yet their (0/2+) transition levels (ε(0/2+)) span a wide energy range (Fig. 4). The ε(0/2+) levels of $V_{O1}$, $V_{O3}$, and $V_{O4}$ lie relatively deep—2.1, 1.8, and 2.0 eV below the CBM, respectively—whereas those of $V_{O2}$, $V_{O5}$, and $V_{O6}$ are shallower—1.4, 1.6, and 1.42 eV below the CBM. As the Fermi level shifts deeper into the band gap, the relatively shallow ε(0/2+) levels of $V_{O2}$, $V_{O5}$, and $V_{O6}$ enable them—despite their initially higher neutral formation energies—to transform into lower-energy charged defects and more readily ionize into the +2 state. Among the six configurations, $V_{O3}$ and $V_{O6}$ exhibit greater stability.

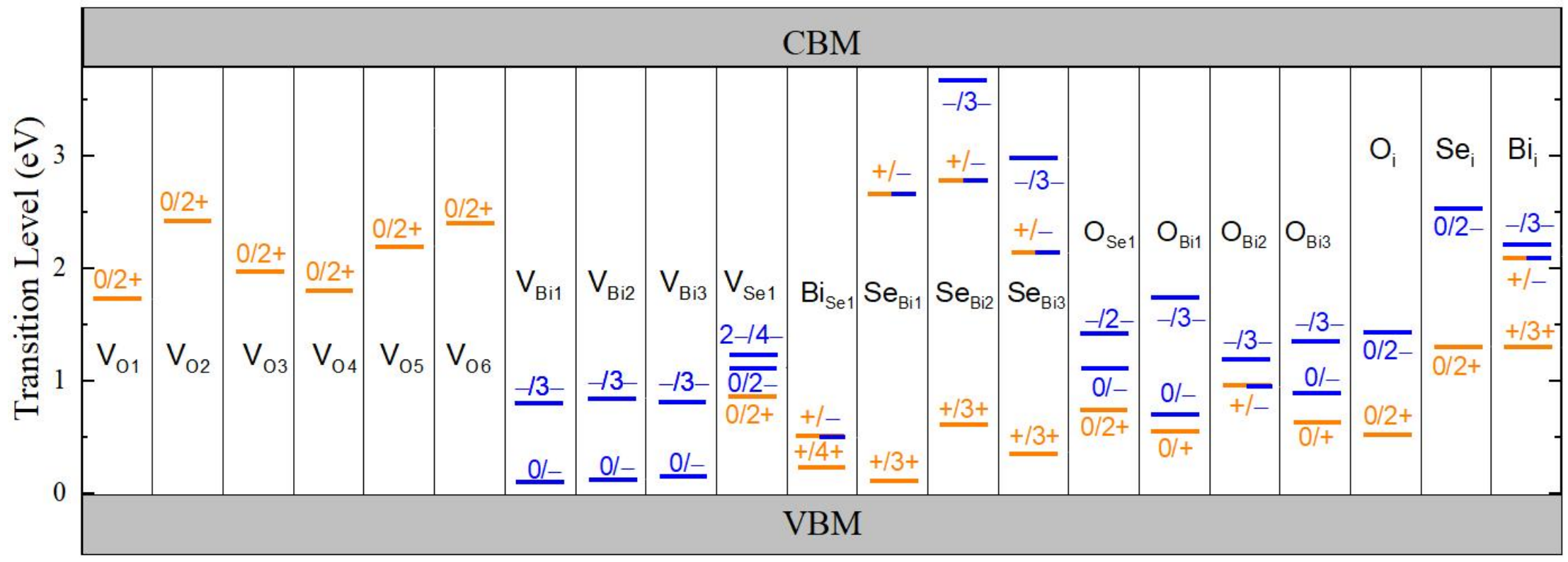

Figure 4. Calculated charge-state transition levels of intrinsic point defects for different defect configurations in the band gap of $Bi_2SeO_5$.

The pronounced variation in formation energies and transition levels across six different $V_O$ configurations highlights the complexity of point defects in the quasi-2D, low-symmetry compound BSO. Each $V_O$ configuration possesses distinct energetics and transition levels, imparting markedly different influences on the electrical conductivity of BSO. Accurately identifying the most stable defects at a given Fermi-level position—and thereby fully capturing the defect physics of BSO—requires a comprehensive study of all configurations and their possible charge states.

Figure 3(c) compares the formation energies and transition levels of $V_{Bi}$ at the three nonequivalent Bi sites with those of $Se_{Bi}$. All $V_{Bi}$ defects act as acceptors: they remain neutral when the Fermi level is close to the valence-band maximum (VBM) and ionize to negatively charged states as the Fermi level shifts upward, thereby generating hole carriers. The minimal differences in formation energies among the three $V_{Bi}$ configurations might suggest that nonequivalent Bi sites could be treated as equivalent. However, our analysis reveals pronounced variations in the formation energies of $Se_{Bi}$ defects, all of which exhibit bipolar behavior—a characteristic attributed to the cationic nature of Se ions in the system. Although the three $V_{Bi}$ configurations show only slight differences in formation energy, the $Se_{Bi}$ defects display strong site-dependent behavior. The $\varepsilon(-1/+1)$ levels of $Se_{Bi1}$, $Se_{Bi2}$ and $Se_{Bi3}$ lie at 2.66 eV, 2.78 eV, and 2.14 eV above the VBM, respectively, whereas their $\varepsilon(+/3+)$ levels appear at 3.69 eV, 3.20 eV, and 3.44 eV below the CBM. These disparities demonstrate that the three Bi sites differ markedly in their tolerance to antisite defect formation, with defect tolerance strongly influenced by changes in charge states.

As depicted in Figure 4, intrinsic defects in BSO—except for $V_{Bi}$— are all deep-level defects. The $\varepsilon(0/2+)$ of $V_O$ lies at least 1.5 eV below the CBM. Both $Bi_{Se}$ and $Se_{Bi}$ ($Se_{Bi1}$, $Se_{Bi2}$, $Se_{Bi3}$) exhibit bipolar behavior, with their $\varepsilon(-/+)$ levels appearing at 0.5 eV, 2.66 eV, 2.78 eV, and 2.14 eV above the VBM, indicating that the charged states are energetically more favorable than the neutral ones. In contrast, all $V_{Bi}$ defects act as shallow acceptors, with their $\varepsilon(-1/0)$ levels of $V_{Bi1}$, $V_{Bi2}$ and $V_{Bi3}$ located just 0.10 eV, 0.12 eV, and 0.14 eV above the VBM, respectively.

## Defect Formation Energies of all intrinsic point defect

Six representative points (A–F) in the stable region as shown in Fig. 2 are selected to calculate the defect properties of synthesized BSO samples. In Figure 5, the formation energies of all point defects in different charge states are plotted as functions of the Fermi level. Solid colored lines denote dominant defects in BSO, whereas gray lines represent defects with relatively high formation energies and consequently lower concentrations. Antisite defects ($Bi_O$ and $Se_O$) and interstitial defects ($Bi_i$ and $Se_i$) exhibit substantially high formation energies, indicating that their concentrations are negligible. According, they are depicted as light-gray lines in Figure 5 without further distinction.

Under the six distinct synthesis conditions, the formation energies of $Se_{Bi}$, $Bi_{Se}$, $V_O$, $O_i$, and $V_{Bi}$ remain uniformly low, identifying them as the dominant intrinsic defects. As shown in Figure 5, across most growth environments, the acceptor defect $Bi_{Se}$ and the donor defect $Se_{Bi}$ have the lowest formation energies. This phenomenon contrasts sharply with substitution defects in conventional compounds such as $Bi_2Se_3$, $Bi_2S_3$, $Sb_2Se_3$ and $Sb_2S_3$[31-35], where substitution defects typically possess very high formation energy. Remarkably, in BSO, both cation-replace-anion antisite defect ($Bi_{Se}$) and the corresponding anion-replace-cation antisite defect ($Se_{Bi}$) emerges as major defects—an unusual phenomenon. In conventional compounds, Se predominantly adopts an anionic role with the -2 valence state; however, in BSO, it exhibits atypical cationic behavior with a +4 valence state. The valence change of Se eliminates its anionic behavior, giving rise to distinct defect properties compared to traditional group-VI compounds. Accordingly, $Bi_{Se}$ and $Se_{Bi}$ should be regarded as cation-replace-cation antisites. This also explains why anion substitution defects ($Se_O$ and $O_{Se}$) have high formation energy in BSO, while substitution defects involving elements from the same main group are more easily formed. As the Fermi level sweeps across the band gap, both $Bi_{Se}$ and $Se_{Bi}$ display amphoteric behavior, acting as acceptors or donors depending on the Fermi level position, which is generally pinned near the intersection of the lowest-energy donor and acceptor formation-energy curves. At this pinning position, $Bi_{Se}$ behaves as an acceptor while $Se_{Bi}$ functions as a donor; their formation energies are lowest, and consequently cation-replace-cation antisite defects predominate in BSO.

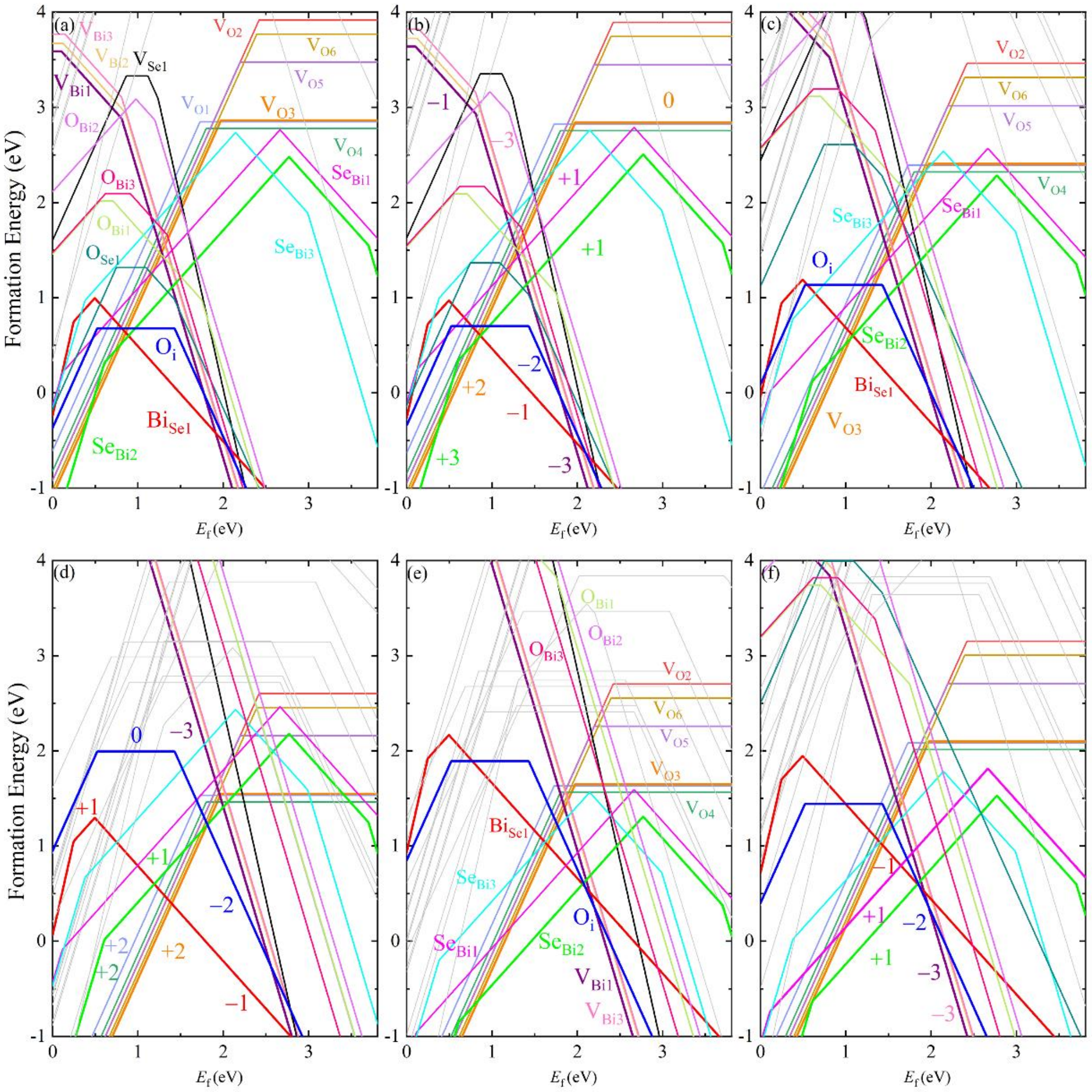

Figure 5. Calculated formation energies of intrinsic point defects in different charge states as a function of the Fermi level under chemical potential conditions (a) A, (b) B, (c) C, (d) D, (e) E, and (f) F shown in Fig. 2. Gray lines denote defects with comparatively high formation energies. For ionized donor defects, the formation energy increases with Fermi level (positive slope), whereas for ionized acceptor defects, it decreases (negative slope). The intersections or turning points of the lines mark the transition energy levels, where the formation energies of different charge states of a defect are equal. The colored “+2” label signifies that the neutral oxygen vacancy $V_O^0$ donates two electrons to become the doubly positively charged defect $V_O^{2+}$.

Among the vacancies, the three nonequivalent $V_{Bi}$ defects are all the shallow acceptors with small transition energy level above the VBM. however, their relatively high formation energies limit their contribution to p-type conduction. The six $V_O$ configurations are deep doors with high formation energy under most conditions; yet, as $\mu_O$ decreases, their formation energies drop rapidly (Firgure 5c-d), making them the dominate donors. The formation energy of $V_{Se}$ is too high to significantly affect the electrical properties. Notably, the neutral formation energy of $O_i$ is lowest among all neutral defects under most growth conditions, consistent with the layered structure of BSO that readily accommodates interstitial atoms.

At the Fermi level pinning position, $V_O$ and $Se_{Bi}$ are most stable in the +2 and +1 charge states, respectively, donating two and one electrons to the lattice and thus acting as the primary donor. Conversely, $Bi_{Se}$, $V_{Bi}$ and $O_i$ are most stable in the –1, –3, and –2 states, each generating one, three, and two holes respectively, and acting as the dominant acceptors. The other intrinsic defects have prohibitively high formation energy, as shown by the gray lines in figure 5, and thus exert negligible influence on the electrical property of BSO. Under most growth conditions, $Bi_{Se}^-$ has the lowest formation energy and acts as effective acceptor, making it a reliable source of p-type conduction. Although $Se_{Bi}$ and $V_O$ are the main donors, their formation energies are generally higher than that of $Bi_{Se}$ under most chemical conditions and they cannot release sufficient electrons to compensate for the holes produced by $Bi_{Se}^-$. Only under Se-rich, Bi-poor conditions (points E and F) do the formation energies of $Se_{Bi}^+$ become comparable to those of $Bi_{Se}^-$, thereby generating more electrons and rendering BSO an intrinsic semiconductor or weak n-type conductor. However, deliberately shifting the Fermi level upward toward the CBM to enhance n-type conductivity proves highly challenging. Because the formation energies of $V_{Bi}^{3-}$ and $O_i^{2-}$ defects decrease sharply, leading to high hole concentrations that counteract any further upward shift of the Fermi level. In fact, their formation energies can even become negative, meaning these defects form spontaneously, compensate free electrons, and pin the Fermi level. Therefore, achieving strong n-type doping in BSO is fundamentally impossible.

As discussed above, no intrinsic defect serves as an effective n-type source (i.e., with shallow transition levels and a low formation energy); therefore, good intrinsic n-type conductivity is unlikely to be achieved. As shown in Figure 5, tuning the chemical potentials shifts the Fermi level, allowing conductivity to vary from relatively strong p-type to weak n-type depending on growth conditions. Under O-rich and Se/Bi-poor conditions (Fig. 5a,b), BSO displays robust p-type behavior due to the facile formation of the $Bi_{Se}^-$ acceptor. In contrast, under O-poor and Se/Bi-rich conditions (Fig. 5e), or at intermediate O, Se, and Bi partial pressures (Fig. 5f), the Fermi level resides at or just above mid-gap, BSO exhibits intrinsic semiconductor characteristics or very weak n-type conduction, arising from the strong defect compensation between $Se_{Bi}$/$V_O$ donors and $Bi_{Se}$ acceptors.

## Conclusion

In summary, we have systematically investigated the defect properties of BSO. Although BSO is chemically ternary compound, Bi and O possess 3 and 5 nonequivalent sites, respectively, which renders its defect chemistry highly complex. Because Se adopts an unusual +4 cationic valence state, the formation energies of main-group anion substitutional defects ($Se_O$ and $O_{Se}$) are very high, and their concentrations can therefore be neglected. $Bi_{Se}$ and $Se_{Bi}$ are the dominate defect, in traditional Bi-Se-based compound, these defects correspond to cation-raplace-anion and anion-replace-cation antisites with very high formation energies; however, they behave as cation-replace-cation antisites in BSO. The variability in the formation energies of the six types of $V_O$ defects indicates that defects of the same type located on nonequivalent atomic sites can exhibit markedly different properties. Under O-rich and Se/Bi-poor conditions, BSO displays relatively robust p-type behavior, whereas under O-poor and Se/Bi-rich conditions, or at intermediate O, Se, and Bi partial pressures, it exhibits intrinsic semiconductor characteristic or very weak n-type conduction due to strong donor-acceptor compensation. This study provides theoretical insights to guide the development of high-performance BSO-based electronic devices.

## ACKNOWLEDGMENTS

This work was supported by the National Natural Science Foundation of China (12404093, 12404089 and 12304110), the China Postdoctoral Science Foundation (2021M702915 and 2022M720813).